# Charge transfer and weak bonding between molecular oxygen and graphene zigzag edges at low temperatures

D. W. Boukhvalov[*,1,2], V.Yu. Osipov[†,3], A.I. Shames[4], K. Takai[5], T. Hayashi[6], T. Enoki[7]

[1]*Department of Chemistry, Hanyang University, 17 Haengdang-dong, Seongdong-gu, Seoul 133-791, Korea*

[2]*Theoretical Physics and Applied Mathematics Department, Ural Federal University, Mira Street 19, 620002 Ekaterinburg, Russia*

[3]*Ioffe Physical-Technical Institute, Polytechnicheskaya 26, 194021, St. Petersburg, Russia*

[4]*Department of Physics, Ben-Gurion University of the Negev, P.O. Box 653, 8410501, Be'er-Sheva, Israel*

[5]*Department of Chemical Science and Technology, Hosei University, 3-7-2, Kajino, Koganei, Tokyo, 184-8584, Japan*

[6]*Faculty of Engineering, Shinshu University, 4-17-1 Wakasato, Nagano 380-8553, Japan*

[7]*Department of Chemistry, Tokyo Institute of Technology, 2-12-1, Ookayama, Meguro-ku, Tokyo 152-8551, Japan*

**Abstract**

Electron paramagnetic resonance (EPR) study of air-physisorbed defective carbon nano-onions evidences in favor of microwave assisted formation of weakly-bound paramagnetic complexes comprising negatively-charged $O_2^-$ ions and edge carbon atoms carrying π-electronic spins. These complexes being located on the graphene edges are stable at low temperatures but irreversibly dissociate at temperatures above 50-60 K. These EPR findings are justified by density functional theory (DFT) calculations demonstrating transfer of an electron from the zigzag edge of graphene-like material to oxygen molecule physisorbed on the graphene sheet edge. This charge transfer causes changing the spin state of the adsorbed oxygen molecule from $S = 1$ to $S = 1/2$ one. DFT calculations show significant changes of adsorption energy of oxygen molecule and robustness of the charge transfer to variations of the graphene-like substrate morphology (flat and corrugated mono- and bi-layered graphene) as well as edges' passivation. The presence of H- and COOH- terminated edge carbon sites with such corrugated substrate morphology allows formation of ZE–$O_2^-$ paramagnetic complexes characterized by small (<50 meV) binding energies and also explains their irreversible dissociation as revealed by EPR.

**Keywords:** nanographene, edge states, oxygen molecule, adsorption, charge transfer, ionic complexes, low temperatures, electron paramagnetic resonance, density functional theory.

---

[*] Corresponding author: E-mail address: danil@hanyang.ac.kr (D.W.Boukhvalov)

[†] E-mail address: osipov@mail.ioffe.ru (V.Yu.Osipov)

# 1. Introduction

Nano-carbons with a developed π-electron system assembled on the basis of nanographene having open edges are within the scope of intent interest due to the promises of such materials for advanced molecular electronics, smart molecular sensors and chemical catalysts [1]. The zigzag-type edges in nanographene possess nonbonding π-electron states (edge state) which have excessive π-electronic charge density located and smeared within the narrow strip with decay length about of ~ 3 aromatic rings in the vicinity of a zigzag boundary of flat nanographene sheet [2, 3]. In contrast, the armchair-type edges do not have the edge state and instead are the source of electron wave interference, which results in the formation of the $\sqrt{3} \times \sqrt{3}$ superlattice of charge density wave. Such edge geometry dependence in the electronic structure is understood on the basis of the edge boundary condition of the massless Dirac fermion running on the two-dimensional honeycomb bipartite lattice. In other words, it can be explained in terms of the migration and localization of aromatic sextets in zigzag and armchair edged nanographenes, respectively, in relation to the Clar's aromatic sextet rule in chemistry [4]. The edge geometry dependence predicted theoretically has been confirmed by scanning tunnel microscopy/spectroscopy and atomic force microscopy studies [5, 6, 7, 8] where the presence of edge state in zigzag edges (ZE) and the formation of the charge density wave in armchair edges have been confirmed.

Interestingly, the edge state localized in a zigzag edge is strongly spin polarized and the edge-state spins are coupled with each other in a zigzag edge through strong ferromagnetic interaction. This makes nanographene be magnetic in spite of the diamagnetic nature of infinite size graphene. In nanographene having arbitrary shaped periphery, which consists of a combination of zigzag and armchair edges, the edge-state spins ferromagnetically coupled with each other within a zigzag edge interact with those in other zigzag edge that is separated from the former one with the presence of a nonmagnetic armchair edge, with the aid of conduction-electron-mediated inter-zigzag-edge interaction, whose strength and sign (ferromagnetic/ antiferromagnetic) vary depending on the mutual geometrical relation. Therefore, the compensation between the ferromagnetic zigzag edges brings about ferrimagnetic structure with non-zero net magnetic moment [9]. The strength and spatial distribution of the edge-state spin depends on what functional groups terminate the edge carbon atoms [8, 10]. Here it should be noted that σ-dangling bond, which is σ-nonbonding state, can be another spin source in addition to the edge state of σ-origin. However, the σ-dangling bonds cannot survive and are terminated by oxygen-containing functional groups such as –COOH, >C=O, -OH in addition to hydrogen [11] when nanographene is handled in the ambient atmosphere. Accordingly, the σ-dangling bond spins are excluded as the spin source.

The large local density of states in the edge state populated in the zigzag edge plays an important role in reactions with foreign chemical species as it is singly occupied nonbonding state

[12]. In fact, nanographene-based nanoporous carbons work effectively as catalyst in various chemical reactions because of their high density of edges, in which chemically active edge states are populated. For example, defective carbon nano-onions (DCNOs) with the specific surface area ~ 400-500 $m^2g^{-1}$ which are produced by high temperature ($\geq 1600$ $^oC$) annealing detonation nanodiamonds with the mean size ~5 nm and narrow particle size distribution (FWHM ~ 1 nm) have one of the highest densities of edges per mass unit [13,14] and, therefore, the highest concentration of localized spins (~$6\times10^{18}$ $g^{-1}$) of the edge states , ever observed among all other members of nano-carbon family except activated carbon fibers [15] and nano-porous zeolite-templated carbon [16]. Actually, paramagnetic properties of DCNO caused by inherent zigzag edge spin sites were well described in details in Ref. 14, 17,18.

Such features of DCNO make them very convenient objects in studying the interaction of guest gaseous agents with edges of nanographene sheets, mainly through especial morphological features of outer shells of DCNO particles which have a lot of split and bended nanographene strips. A lot of studies in the past have been devoted to analysis of gases adsorption/ desorption by various carbonaceous and graphenic-like materials, including oxygen adsorption/ desorption from activated carbon fibers, carbon nanotubes and graphene [19-22]. At the same time a little attention has been paid to gases adsorption on the DCNO itself. This is because basic studies of DCNO have been intensively started just around 2000 after the great progress happened in the technology of post-fabrication extraction and purification its precursor material – detonation nanodiamonds, the new carbon material appeared in the global market.

Physisorption of oxygen molecules on the basal planes of graphene or highly oriented pyrolitic graphite has been investigated in details during last four decades both experimentally and theoretically. The weak binding of triplet oxygen molecule with graphene plane is provided by van-der-Waals forces with small binding energy (~0.1-0.12 eV). The distance between $O_2$ molecular axis and basal plane equals ~0.30 nm, but no more than ~0.37 nm in this case [23, 24]. The amount of charge transfer from the graphene to the physisorbed $O_2$ molecule is negligible small and about of ~0.03$e$ per molecule [24]. The edges sites of graphene/graphitic planes are much more chemically active than sites on the basal sheets and easy form covalent chemical bonds with foreign species [23]. Following to the theoretical predictions done in Ref. 23 chemical covalent bonding of oxygen molecule with zigzag edge carbon atom (C-O distance equals 0.165 nm) leads to the *quenching of local magnetic moments* of this underlaid edge carbon atom and four nearest neighboring carbon atoms. Oxygen molecule can even dissociate near the edge and form two strong >C=O carbonyl bonds [21] with two outermost edge carbon atoms (chemisorption). Other defective pieces of graphene (vacancies, holes, dislocations) are also chemically active and participate in covalent chemical bonding with oxygen. Nevertheless, the intermediate complexes with binding energies between the van-der-Waals and strong covalent bonding were not so much investigated. Taking into account that DCNO has a great concentration of edges and edge-localized spin states it is much promising to study their interaction with molecular oxygen as a guest paramagnetic agent [18]. Because both agents are paramagnetic it is possibly to apply magnetic techniques for studying their interaction in the whole temperature range from liquid helium temperature up to room temperature.

Recently we unexpectedly found that zigzag edge in DCNO of quasi-spherical shape[2] which has loose graphene quality of exterior outermost carbon shells and contains adsorbed ambient air gases including molecular oxygen, in appropriate microwaves-assisted conditions at cryogenic low temperatures ($T < 50$ K) forms unconventional spin-half $O_2^-$ species. We investigated the interaction of molecular oxygen with π- electronic edge-localized spins of DCNO particles by means of EPR method and found the appearance of new stable EPR-visible magnetic molecular complexes zigzag edge (ZE) – oxygen ion-radical $O_2^-$ at temperatures below 50 K [25]. Such new complexes never observed before can be easy generated at temperatures below 15 K by quick overheating and mutual charging[3] of nanoparticles in the conductive air-physisorbed DCNO powder sample by X-band microwave radiation (200 mW) of EPR spectrometer and its subsequent quick cooling back to the main temperature of cryostat after switching-off the incident microwave power or its reduction down to ~0.1 mW. It was found in Ref. [25] that new EPR-visible complexes ZE–$O_2^-$ are very stable at low temperatures below ~20 K, persist for hours (at least) and irreversibly destroys when once heating above 50-60 K. These phenomenological results suggest a presence of quite interesting spin magnetism in this system. However, the microscopic origin of these weakly-bound paramagnetic complexes on the edges has been still unclear. So, a detailed examination of the experimental results obtained by using more well-defined sample and theoretical analysis of the phenomena are requited. For making DCNO particles with more perfect and well-defined structure of honeycomb nanographene lattice the use of heat treatment at more higher temperatures ( >1800 °C) is highly desirable. According Raman spectroscopy studies the increase in heat treatment temperature leads to substantial reduction the number of pentagon-heptagon carbon ring clusters in the outermost shells of DCNO and improvement their graphene quality [26]. In this paper we *specially investigated* the appearance of weakly-bound complexes ZE- $O_2^-$ in DCNO with well-resolved graphene structure obtained at ~1950 °C.Density functional theory (DFT) is widely used method for the study of electronic and magnetic structure of graphene edges [27] and adsorption of oxygen on graphene. Previous studies usually discussed chemisorption of molecular oxygen in the vicinity of graphene edges and discussed further oxidation and unzipping of graphene [28] and possible catalytic effects of pure and doped graphene [29]. The formation of unusual ionic bonds of molecular oxygen and the edges of graphene nanosheets was discussed in the Ref. [23]. Previously studied differences between chemical properties of graphene nanosheets and quasi-infinite graphene [30] demonstrate differences in chemical properties between them. Our previous works (see for review [31]) demonstrate that correct description of chemical properties of graphene is required taking into account all possible realistic conditions.

---

[2] Such DCNO of quasi-spherical shape with various aspect ratios varying between 1:1 and ~1:3 were fabricated by annealing well purified detonation nanodiamonds at 1600 °C.

[3] It happens through microwave eddy currents circulating at high strength of electric field vector.

In this work we also performed modeling based on density functional theory calculations of the adsorption of molecular oxygen on zigzag edges of flat and corrugated mono and bilayer graphene and also take into account possibility of passivation of graphene edges by various functional groups (-H, -OH, -COOH) and co-adsorption on the edges other molecules present in the air ($N_2$, $CO_2$ and $H_2O$). By means of these theoretical modeling, we investigated the microscopic view of coupling between oxygen and edge carbon sites responsible for appearance of the unusual spin phenomenon in air-physisorbed DCNO.

## 2. Experimental

The DCNO particles were obtained by high temperature ($T = 1950\ ^oC$) heat treatment of acid-purified detonation nanodiamond particles (mean crystal size ~5 nm, ash content < 0.03 wt.%) in the Ar atmosphere, i.e. by means of direct diamond-to-graphite phase conversion. Trace analysis of precursor detonation nanodiamond powder shows about 9 ppm of Fe, 2 ppm of Mn, 1 ppm of Ni, 70 ppm of Ti, ~100 ppm of B, the content of other residual metals was less than 0.2 wt. ppm level for each pollutant. The details of annealing process were discussed by us earlier in Ref.14. At $T \geq 1600\ ^oC$ almost all nanodiamond particles converted into DCNO particles. However, here, the exploited heat-treatment temperature is much higher than that for the sample in the previous report (1600 $^oC$) [25], resulting in the sample having well-defined graphene sheet and edge structures, which are necessary for detailed examination of experimental results to compare with theoretical results for the ideal structures.

High resolution transmission electron microscopy (HRTEM) (JEOL JEM-2100F equipped with CEOS Cs correctors, 80 kV) was used to investigate the structure of obtained materials as a result of heat treatment at $T = 1950\ ^oC$. Other structural characterization of this material was done by Raman spectroscopy in Ref.[26].

Magnetic susceptibility measurements were done by means of SQUID magnetometer MPMS-7 (Quantum Design, US) at fixed magnetic field 1 T in the whole temperature range 2-300 K. About ~20 mg of DCNO sample was packed in the aluminum foil and disposed in the middle of long (~200 mm) glass tube filled by air at pressure ~15 mbar or well evacuated in case of physisorbed-gases-free sample. The later sample was sealed in the glass tube after prolonged heat treatment at 450 $^oC$ at pressure $\leq 1\times10^{-5}$ mbar and then measured by SQUID magnetometer in advance of subsequent filling this tube by air at low pressure. Temperature dependence of magnetization of molecular oxygen adsorbed on DCNO particles was obtained as a difference between magnetic susceptibilities of air-physisorbed and physisorbed-gases-free samples. This is because oxygen is the only magnetic agent in air presented in content of 23 wt.%. In terms of magnetic susceptibility values the difference between magnetizations of both oxygen-containing and oxygen-free samples is normalized by weight of carbon host sample and then divided by corresponding magnetic field value expressed in Gauss units.

EPR measurements in the temperature range 4 K ≤ $T$ ≤ 90 K were carried out using a JEOL JES-TE200 X-band (□ ~9 GHz) EPR spectrometer equipped with Oxford Instrument ESR900 cryostat at microwave power $P_{MW}$ ranging from 50 µW to 200 mW. For EPR study the loose packed sample (~4 mg) was sealed in 1 mm internal diameter (i.d.) ~45 mm long quartz capillary tube at ambient conditions (air pressure 1 bar). The sample under study is designated as HT1950-air. The whole content of gaseous agents (including water vapor) adsorbed on the DCNO at room temperature from air was about ~ 5 wt.%. The weight of gaseous air in the sealed capillary does not exceed ~0.05 mg that corresponds to about ten oxygen molecules per single DCNO particle consisting of ~$10^4$ carbon atoms. Processing and simulation of EPR spectra were done using JEOL-JES-TE200 and Microsoft Excel software. Temperature dependences of doubly integrated intensity $DI \sim I_{pp} \times ( \Box H_{pp})^2$ ($DI$ is proportional to the EPR susceptibility $\chi_{EPR}$) were analyzed. Here $I_{pp}$ and $\Box H_{pp}$ are peak intensity and peak-to-peak line width of the EPR signal recorded in the conventional form of 1$^{st}$ derivative of EPR absorption spectrum.

## 3. Computational method

The modeling of the adsorption of oxygen and other molecules on the edges of graphene is carried out by the density functional theory realized in the pseudopotential code SIESTA. [32] All calculations are done using the local density approximation (LDA). [33] This approach was used in our previous studies of adsorption of molecules on graphene and charge transfer in these systems [34-36]. All calculations were carried out for energy mesh cut off 360 Ry and k-point mesh 4×4×2 in the Mokhorst–Pack scheme. [37] Number of k-point along z direction perpendicular to graphene plane is 2 because graphene with ripples is not strictly a two dimensional system. During the optimization, the electronic ground state was found self-consistently using norm-conserving pseudopotentials [38] for cores and a double-ζ plus polarization basis of localized orbitals for carbon, oxygen and nitrogen and a double-ζ basis for hydrogen. Optimization of the bond lengths and total energies was performed with an accuracy of 0.04 eV/Å and 1 meV, respectively.

For the modeling of the graphene edges were used graphene nanoribbons of 5 nm width within periodic conditions along ribbon separated by 2 nm of empty space along z axis. For test of validity of used model we perform the calculations for the nanoribbons of 4, 3 and 2 nm width and also vary the size of supercell along ribbon from 0.5 to 2 nm. We considered zigzag edges functionalized by atomic hydrogen, hydroxyl (-OH) and carboxyl groups (-COOH) and used narrower nanoribbons for reduction of computational costs to explore the maximal load of oxygen and other molecules present in the air ($N_2$, $CO_2$ and $H_2O$) on flat graphene edges. For all discussed width of the graphene nanoribbons the values of migration energies, sizes of the areas of charge harvesting (see below) and charge transfer were found almost the same (deviation of the values of energies is a few meV per system).

For the modeling of realistic DCNO which have deviation from planar geometry we also used corrugated graphene bilayers. To build the model of the corrugation all atoms in the centre of the supercell have been smoothly shifted and several atoms on the top are fixed. Several atoms on the edges of the supercell are also fixed on the initial (zero) height of the graphene flat. Further optimization of the atomic structure provides formation of the smooth surface of corrugated graphene similar to what was obtained for the ripples in infinite graphene sheets. This method is the same as used in our previous works for the modeling of extrinsic corrugation of graphene flat [36, 39].

The value of charge transfer between graphene and adsorbed molecules is defined as the difference between the total number of electrons on the valence orbital of the studied molecule and calculated occupancies of the same orbitals after adsorption. The binding energy is defined as the difference between the total energies of two configurations: oxygen molecule forms bond with carbon atom of basal plane in vicinity of the edge and with utmost carbon atom of the edge. To estimate the temperature of desorption from the edges of corrugated bilayer graphene we performed the test calculations for the carbon system with known exact atomic structure and experimentally measured temperature of oxygen desorption. Graphite surface is satisfying both of these two conditions. For the modeling of graphite surface we used 6 layered slab with 32 carbon atoms in each layers.

The calculation of exchange interaction between distant oxygen ion-radical and edge localized spins was performed for Heisenberg type exchanges with Hamiltonian $H_{ij} = J_{ij}S_iS_j$ by the formula: $J_{ij} = (E_{FM} - E_{AFM})/S_iS_j$, where $E_{FM}$ and $E_{AFM}$ are the total energies of studied system with parallel and antiparallel orientation of spins and $S_{i,j}$ are the values of spins on magnetic centers.

## 4. Experimental results

Typical HRTEM image of the ensemble of DCNO particles obtained at 1950 °C is shown in Fig. 1a. The optimal lateral resolution used allowed distinguishing the graphene interlayer spacing and internal structure of isolated DCNO particles as well as fine honeycomb structure of isolated large-scale graphene flakes appeared in the DCNO powder in minor quantity. The images demonstrate that mean in-plane size of graphitic crystallites equals about few nanometers, and the amount of disordered $sp^2$-phase originating from the presence of five- and seven- membered carbon ring clusters in this sample is much smaller in comparison with that one synthesized at lower heat treatment temperature ($T \approx 1600$ °C) reported recently in Ref. [25]. The high temperature (1950 °C) sample being compared with that synthesized at $T \approx 1600$ °C shows that particles become much more polygonized and have well-graphitezed flat facets (Fig.1a). This is because at temperatures exceeding ~1800 °C carbon atoms can diffuse and jump more intensively within the graphene layers and quickly repair the defects such as vacancies and multi-vacancies in the lattice in the course of annealing process [40]. Well-formed honeycomb graphene structure was easily resolved, for example, on the upper facets of such DCNO particles (Fig. 1a). Other than round-shaped or polygon-like DCNO particles, some isolated bi-layered and few-layered flat graphene sheets with rough edges of arbitrary shape were found in the powder of obtained sample (Fig.1b). Some graphene sheets demonstrate edges with well-resolved zigzag chains with length up to 8-10 zigzags. Regular structure of these large-scale graphene sheets (> 30 nm) overmodulated by moiré pattern clearly visible in Fig.1b confirms that practically similar quality nanographene sheets/ strips were formed in DCNO particles. The outer shells of DCNO particles are defective and split even in the DCNO sample obtained by heat-treatment at 1950 °C because they come from the external rough surface layers of nanodiamond particles which can be represented as consisting of badly terraced facets (111) of diamond lattice and also because the surface of precursor nanodiamond is terminated by oxygen-containing carboxyl and hydroxyl groups [41].

The temperature dependence of magnetization[†] of molecular oxygen adsorbed on the surface of DCNO particles expressed in magnetic susceptibility units is shown in Fig. 2. The actual magnetization of adsorbed oxygen per square unit of adsorbing surface may be easily obtained by dividing the experimental data by specific surface of DCNO particles (~450 $m^2g^{-1}$) and multiplying by magnetic field value. The magnetization levels near the zero value within the temperature range 140-300 K. The increase in magnetization on lowering temperature below 140 K is related with acceleration of the oxygen physisorption on the DCNO surface and substantial increase in number of oxygen molecules physisorbed on the surface. At $T$ ~65 K the magnitization of adsorbed oxygen molecules reaches maximum and then turns down to vanishing value within the temperature range from ~65 to ~30 K. It means that below ~65 K oxygen molecules adsorbed and thereafter immobilized on the DCNO surface and inside the nanopores of aggregates of DCNO particles interact antiferromagnetically and probably form 2D and 3D clusters with total zero spin. Subsequent increase in magnetic moment on further lowering the temperature below ~30 K is caused by Curie paramagnetism of remaining isolated excessive triplet oxygen molecules which do not subjected the simple pairing with neighboring oxygen molecules or alternatively the small amount of appeared exchange-coupled trimers of $O_2$ molecules. Importantly, antiferromagnetic spin pairing between $S=1$ physisorbed oxygen molecules in 2D and 3D clusters becomes substantial just at $T < 50$-$55$ K when the clusters of oxygen molecules immobilized and solidified on the surface and inside the nanopores of DCNO aggregates go to the non-magnetic state [42]. Similar observation done within the same temperature range was reported by Kawamura at al. for oxygen adsorption on activated carbon fibers [43]. Additionally we found that difference between magnetic susceptibilities of air-physisorbed and physisorbed-gases-free samples ($\Delta\chi$) is proportional to the specific surface of nanoparticles. The maximum for $\Delta\chi$ at ~65 K drops ~5 times after replacement the DCNO powder by the powder of nanographite particles having mean size ca 30 nm. For the micron-sized graphite powder the maximum in $\Delta\chi$ at ~60-65 K is found to be about one order of magnitude smaller than that for DCNO. It therefore confirms that data presented in Fig.2 are related rather to physisorption of molecular oxygen on the surface of DCNO particles and inside the nanopores of DCNO aggregates than to oxygen condensation in aluminum foil container.

---

[†] Here we consider the magnetic moment of adsorbed oxygen molecules divided by magnetic field and normalized on the weight of carbon host matrix.

In this study EPR measurements on the HT1950-air sample were done in the same manner as described in [25]. Room temperature spectrum (not shown) demonstrates broad ($\Delta H_{pp}$ = 1.75 mT) Lorentzian-like EPR line with $g = 2.0012$−$2.0014$ attributed to π- electronic edge state spins [14, 18, 25]. On cooling the edge spins' EPR line broadens and loses its peak intensity becoming unobservable (below background level) at $T$ approaching 140 K. On disappearance of the edge spins' line another weak narrow EPR line with $g = 2.0022(2)$ and line width $\Delta H_{pp} = 0.39(1)$ mT becomes observable within the entire low temperature range. Following [25] cooling to low temperatures was done at very low levels of incident microwave power $0 < P_{MW} < 100$ μW. Red trace in Fig. 3a shows EPR spectrum of the HT1950-air sample recorded at $P_{MW} = 100$ μW and $T = 8.5$ K. Then, keeping the same $T = 8.5$ K, the sample was undergone to the microwave irradiation at the higher power level $P_{MW} = 200$ mW during 10 min. Then the power level was abruptly reduced back to $P_{MW} = 100$ μW. Blue trace in Fig. 3(a) shows EPR signal recorded at low ($P_{MW} = 100$ μW) microwave power level 15 minutes after abrupt reduction of the power level. It is clearly seen that the intensity of the same narrow EPR signal with $g = 2.0022$ and $\Delta H_{pp} = 0.39$ mT increases about six times. This new enhanced EPR line was attributed to the new paramagnetic ZE–$O_2^-$ complex [25]. It has the $g = 2.0022$ and its line width is insensitive to adsorbed triplet oxygen which had been responsible for the broadening and disappearance of the edge spins' EPR line at low temperatures. These differences allow easily distinguishing the EPR-observable ZE–$O_2^-$ complexes with $S = 1$ from the underlying zigzag edge-localized spin states with the half-integer spin which in fact become absolutely EPR-unobservable at $T < 140$ K. This new EPR signal disappears at ~ 60 K. Thus, at $T$ above 60 K the EPR signals recorded at the same low power level are the same (even in fine details) disregarding if the sample has passed the high power irradiation at low temperature or not - see Fig.3(b).

Following Ref. [25] the process of microwave generation of the ZE-$O_2^-$ species may be described as follows: (a) local microwave heating causes fast desorption of molecular oxygen (being mostly in energetically favorable triplet state) from facets and edges of DNCO particle; (b) on abrupt switching off high MW irradiation, due to low temperature of the sample kept by the cryostat, triplet oxygen molecules fast adsorb on DNCO. This adsorption process significantly differs from the adsorption occurring during continuous slow lowering temperature. In result of such an adsorption, part of triplet oxygen molecules accept additional electrons from edges and turn to superoxide ion-radicals $O_2^-$ having $S = 1/2$. (c) Then part of superoxide ion-radicals interacts with some neighboring π- electronic edge spins having also $S = 1/2$ creating exchange coupled ZE-$O_2^-$ complexes having $S = 1$ with exchange integral $J \sim -20$ K and dissociation energy ~ 50 K.

The doubly integrated intensity ($DI \sim \chi_{EPR}$) of the new ZE–$O_2^-$ signal recorded at $P_{MW}$ ~100 μW demonstrates quite unusual temperature behavior upon heating (Fig. 4, curve 2). Below 10 K the spin susceptibility levels or slightly decreases evidencing presence of antiferromagnetic exchange interaction between constituents of the ZE–$O_2^-$ complex. Above 10 K $DI$ smoothly drops on raising $T$ following to the non-Curie law in the range 10-45 K, and then, above 45 K, abruptly drops to an initial small vanishing value at ~60 K. It indicates through irreversible decay of ZE–$O_2^-$ magnetic complexes [25]. Trace 1 in Fig.4 corresponds to the temperature behavior of $\chi_{EPR}$ of the same HT1950-air sample in the initial state, i.e. when new magnetic complexes ZE–$O_2^-$ were not still generated. The corresponded temperature behavior (trace 1) approximately follows Curie law ($\sim T^{-1}$) at $T > 20$ K and means that the density of paramagnetic defects responsible for that narrow EPR signal ($g$=2.0022) is constant within the wide temperature range up to 150 K, but much smaller than the density of novel microwave-generated ZE–$O_2^-$ magnetic complexes having practically the same $g$-factor and line width as the narrow EPR signal existing prior the intensive microwave irradiation. It seems that both narrow EPR signals (before and after mw irradiation) have the same microscopic origin, but the new enhanced EPR signal, which appears after the intensive microwave irradiation, originates from complexes with low binding energy and, as a result, lower dissociation temperature. The schematic microscopic model of ZE–$O_2^-$ paramagnetic complexes firstly proposed in Ref. [25] is shown in *Graphical Preview*. It explains how an exterior site of the zigzag edge donates one π-electron to triplet oxygen molecule appeared near the edge and after such accepting the neutral oxygen molecule turns to an ion-radical with spin ½ and charge -1(e) interacting magnetically with π-electronic edge-localized spin. By this way the oxygen anion $O_2^-$ forms more or less weak ionic bond with underplayed zigzag edge of flat or bended nanographene sheet until the thermally activated back charge transfer from $O_2^-$ to zigzag edge will suddenly happen at elevating temperature. Further theoretical consideration based on density functional theory method done in the next section should elucidate the microscopic reasons providing the appearance of such very easy dissociated (above 50 K) paramagnetic complexes.

## 5. Computational results and discussion

### 5.1. Mechanism of charge transfer

The first step of our modeling is the description of changes in charge distribution in the vicinity of zigzag edges of flat graphene passivated by single hydrogen atoms before (Fig. 5a) and after (Fig. 5b) adsorption of molecular oxygen on the edge carbon atom. Physisorption of molecular oxygen in the basal plane part about edges (Fig. 5a) increases the Mulliken population (estimating partial atomic charges) of $O_2$ molecule from 12.000 $e^-$ to 12.199 $e^-$ and decreases magnetic moment from 2.00 to 1.57 $\mu_B$. Note the exact coincidence of the Mulliken population of single oxygen molecule in gaseous phase with the number of valence electrons demonstrates exactness of calculation of this value with used technical parameters such as radii and basis sets. This charge was collected almost homogeneously from all first, second and third neighbors of the carbon atom connected with oxygen. Generally the distribution of charge and magnetic moments in the vicinity of zigzag edge is rather close to that before physisorption of $O_2$. The utmost carbon atoms on zigzag edge collect the charge of about 0.06 $e^-$ per atom from carbon atoms of the second and third rows from the edge (for more detailed numerical information see Fig. S1a). Physisorption on basal plane of graphene provides only a shift of the peak in the density of states (Fig. 5a) but the electronic structure remains rather similar with the electronic structure of oxygen molecule in an empty box.

Adsorption of molecular oxygen over one of the utmost carbon atoms of the same edge of flat graphene (Fig, 5b see also S1b) provides more significant increasing of the Mulliken population of the oxygen molecule (about 0.8 $e^-$) that drastically changes the electronic structure of oxygen (Fig. 5e) and changes the total moment of oxygen molecule to 0.88 $\mu_B$ that is corresponding with the change of spin of the molecule from 1 to ½ (see Fig. 7) and decreases the carbon-oxygen distance from 2.6 to 1.5 Å. This distance is longer than usual value for C-O bonds (1.2-1.3 Å) that evidences for the absence of covalent bond formation. This significant charge transfer makes the bond between graphene and $O_2$ rather robust (see Table. I) and adsorption of oxygen on the edge can be discussed in terms of ionic bond. The electronic structure of oxygen turns from free molecule-like collection of well separated energy levels to solid state-like continuous bands (Fig. 5b) that also evidences formation of strong bond with carbon substrate. Transfer of the non-integer number of electron corresponds to formation of the sharp peak in the vicinity of the Fermi level typical for the case of distribution of electron between a few atoms (Fig. 5b). The charge transferred to $O_2$ is collected inhomogeneously (for more detailed numerical information see Fig. S1b). Similarly to the chemisorption of the hydrogen atom on basal plane of graphene [28] we observe alteration of the atoms with increased and decreased Mulliken population (see Fig. 6a). This effect appearing after adsorption and caused by inequivalency of graphene sublattices. The area of charge harvestings (see inset on Fig. 6a) is determined by the initial distribution of unpaired π- electrons from the edge carbon atoms [44]. The charge redistribution on carbon substrate caused by formation of discussed bond with oxygen provides also colossal changes of magnetic properties of zigzag edges and turns them from magnetic to almost nonmagnetic in vicinity of adsorbed oxygen molecule (Fig. 6b). This effect can be explained by the violation of localization of unpaired π- electrons on the zigzag edges. [1, 14] This vanishing of the magnetism on the edges can be the key to the understanding of the environment related effects of the magnetism of the edges in graphitic systems. [45, 46]

This effect could be interesting for the case of narrowest graphene nanoribbons and smallest graphene nanosheets and understanding the role of the interactions between edges that play no role for the edge effects of larger samples. Increasing of the supercell size along edge is corresponding with increasing of the distance between adsorbed oxygen atoms and does not make any valuable effect for oxygen adsorption. But decreasing of the lateral distance between adsorbed oxygen molecules provides the overlap of the areas of charge harvesting (see inset on Fig. 6a). Adsorption of the oxygen molecule with charge transfer collects the electrons from the carbon substrate and makes next edge atom unusable for the charge transfer because a part of the charge used for the ionic bond formation has been already collected. Only physisorption can be realized on this nearest neighboring site and only next-next carbon site suitable for the formation of ionic bond with molecular oxygen (Fig. 6b). The values of charge transfer and binding energy in this case will be almost the same as in the case of absence of additional physisorbed oxygen atoms (Tab. I and II).

For the estimation of the area, in which the values of magnetic moments decreases and understanding the limits in oxygen adsorption with formation of ionic bonds we perform the calculations for supercell with large lateral size for avoid the crossing of the areas where Mulikken populations on atoms is visible changed after adsorption of oxygen and calculate diminishment of moments and changes of electronic structure of utmost carbon atoms (Fig.8b). The results of calculations demonstrate a significant (more that 80%) diminishment of magnetic moments on the nearest utmost carbon atoms that also corresponding with significant decay of the Mulliken populations and smooth decay the values of magnetic moments with increasing distance from adsorbed oxygen. Electronic structures of carbon atoms also demonstrate smooth changes and on the borders of demagnetized area are almost the same as in zig-zag graphene edges. (Fig. 6) For the estimation of magnetic exchange interaction between magnetic oxygen molecule adsorbed on graphene edge and magnetic carbons atoms outside demagnetized area we perform the calculations of total energies for the cases of parallel and antiparallel orientation of spins on oxygen and carbon atoms and find that antiferromagnetic configuration estimated is about 6 meV energetically favorable.

## 5.2. Effect of carbon substrate shape and edge functionalization

The binding energy between the edge of flat graphene monolayer passivated by single hydrogen atoms and oxygen molecule is 405 meV. This value is rather close to the energy required for water evaporation (430 meV) [47] which are corresponding to stability of studied C-O ionic bond to the room temperature that is much higher than experimentally obtained temperatures of oxygen desorption (50-60K). To make our modeling more realistic we performed calculations for different functional groups on the edges that can be on the edges of nano-onions. Of course, the defective nano-onions playing the role of carbon substrates in our experiments are not true flat monolayers with infinite zigzag edges, although some split, twisted and bended nanographene strips occurs on the outermost shells of nano-onions and isolated few-layered graphene sheets (size < 100 nm) also appear in the powder of nano-onions. To take this into account we also perform the calculations for the graphene bilayer which have chemical properties closer to few layer than to monolayer [48] and could be a good model for the description of the adsorption on various types of graphene multilayers and its edges. Our previous calculations [36] suggest that the changes of binding energy and charge transfer can be changed by corrugation of graphene substrate. We also take into account this effect and perform calculations for the maximal ratio of height to width of corrugated area (see Fig. 5c). This value (0.16) is corresponding with the significant corrugation of graphene sheet. [39]

The results of calculations (Tab. I) demonstrate that functionalization of edges by hydroxyl groups (-OH) enhances the binding between graphene edges and oxygen molecules because this group can supply more electrons for the formation of ionic bonds. Carboxyl groups on the edges (-COOH) provide opposite effects because they pull electrons from the edges. The presence of the second layer provides moderate effect on the charge transfer and binding energy in contrast to corrugation of graphene sheet that valuable decrease the charge transfer and binding energy because a part of the charge moves from the edges to the second layer or to the central part graphene nanoribbons [36, 39]. Mulliken population on carbon atoms in vicinity of the edge varies at the same atoms independently from the morphology of carbon substrate (for more detailed numerical information see Fig. S1). Varying the shape of graphene and chemical species on the edges provides only qualitative changes in the populations and magnetic moments of the carbon atoms but the area of the electron harvesting remains the same, because the general pattern of the changes in the Mulliken populations and magnetic moments is determined by the general properties of graphene (hexagonal lattice, two sublattices of carbon atoms) and shape of the edge (zigzag).

These changes of the morphology of carbon substrate provide changes of electronic structure of adsorbed oxygen molecule (see Fig. 5c) which becomes closer to electronic structure of oxygen molecule physisorbed over the basal plane part of graphene. Decreasing of the value of electrons transferred from carbon substrate to oxygen molecule and binding energy provide the increasing of carbon-oxygen distance and additional rise of second oxygen atom (Tab. II). Thus we can say that weakening of the graphene-oxygen bonds is corresponding with slight up-standing of oxygen molecule from quasi-planar position to vertical direction but projection of oxygen molecule on graphene plane remain the same – perpendicular to the edge (Fig. 5b,c). The binding energy between edges of corrugated bilayer and oxygen molecules is about ten times smaller than for flat graphene monolayer (46 and 405 meV respectively, see Tab. I) because the Mulliken population of atoms in equivalent positions in the vicinity of edges is smaller by the amount of 0.010~0.002 $e^-$ per atom than for the flat graphene (for more detailed numerical information see Fig. S1). The total magnetic moment of the adsorbed oxygen molecule is higher than that in the case of flat substrate - 0.96 vs 0.88 $\mu_B$. The nature of the redistribution of the charge in corrugated graphene is the same as previously discussed for the bended graphene [49]. These apparent negligible diminishments of the Mulliken populations of the carbon atoms provide a decrease of the total value of charge transfer by the value of 0.02 $e^-$ (see. Tab. I). In the case of graphene bilayer described above the vanishing of magnetic moments on the edges occurs only on top layer connected with oxygen, magnetic moments on the bottom layers survive.

The calculated binding energy between oxygen molecule physisorbed on graphite is 26 meV which can be compared with experimental value 40 K [50]. Thus the slightly higher binding energy order of 40 meV could be corresponding with bigger desorption temperatures about 60~80 K. Despite described above, the decrease of charge transfer and carbon-oxygen distance and colossal (almost ten times) decrease of binding energy and magnetic moment of oxygen molecule remain almost the same and variation of type and morphology of carbon substrate and functional groups on the edges play the role only for thermal stability of the discussed systems because the presence of unpaired electron on zigzag edges is rather energetically unfavorable that makes these edges extremely chemically active. [2] When deliverance from unpaired electrons cannot be realized by oxidation it can be obtained by transfer of "excess" electron to oxygen molecule. Varying of the shape of carbon substrate and chemical groups on the edges does not provide vanishing of unpaired electron on the edge and could play a role only in for the value of charge transfer but not for the existence of this process.

We also perform the check of alternative reasonable from chemical point of view configurations which can be the source of magnetic moment with S=1/2 on oxygen atoms: >C*-O-O-O*. It happens when oxygen molecule bounds with oxygen atom of edge carbonyl group >C=O. But this configuration is extremely unstable and turns back to >C=O + $O_2$ (with S=1 on oxygen molecule) because this structure contains no dangling bonds. Thus we can conclude that all configurations discussed above are unique that can provide S=1/2 on oxygen atom bonded with DCNO.

## 5.3. Effect of co-adsorption of other gases

In the experiment adsorption of oxygen molecules occurs from the air and other molecules of oxygen, nitrogen, carbon dioxide and water also can be adsorbed in the vicinity of studied bond. For the check of the effect of co-adsorption of other molecules from the air we study the cases of adsorption of the second molecules in the vicinity of already adsorbed oxygen. The results of our calculations (Fig. 7 and Tab. II) give evidence of negligible effect of other oxygen molecules and valuable more than twice increase of binding energy and charge transfer when nitrogen and water molecules physisorbed on the nearest edge carbon atoms. The nature of this effect is the charge transfer from these foreign molecules to the edges (about 0.10 $e^-$) that supply more electrons to substrate and larger back transfer of charge from carbon to oxygen. Further increasing of the concentration of nitrogen provides simultaneous increasing of charge transfer, binding energy and decreasing of magnetic moment on oxygen molecule to the value 0.812 $\mu_B$. The cause of this decreasing of magnetic moments is the tendency to filling last unoccupied $2p$ orbitals of oxygen. Note that this situation could occur only at the infinite flat zigzag edge but in the studied systems this type of edges has limited length and rather rare and discussed situation are almost impossible. Experimental results suggest than carbon dioxide also can be ionized by transferring the electron from outside [19] and possible also can harvest unpaired electrons from graphene edges. In contrast to water and nitrogen carbon dioxide pull the charge (about 0.15 $e^-$) from the graphene edges that provide decreasing of charge transfer and binding energy but the value of charge transferred from carbon substrate is not enough for the formation of ionic bonds and creation of visible magnetic moment on $CO_2$ molecule. Comparison of the effect of carbon substrate shape (Table I) and co-doping (Table II) demonstrate prevailed effect of substrate shape and edges functionalization rather than effects caused by co-doping.

For the estimation of the temperature ranges of co-adsorption we performed the calculation of binding energy of co-adsorbed gases and find that these energies vary from 14 to 44 meV (Tab. II) that are always smaller than the carbon-$O_2$ binding energies. Based of our estimation of relation between calculated binding energies and known desorption temperatures we can conclude that effect of co-adsorption is valuable at temperatures below 50 K. At higher temperatures molecules of co-adsorbents will be desorbed. Thus we can conclude that other molecules from the air play no role in studied phenomenon.

## 6. Conclusion

Low temperature ($T < 45\text{-}50$ K) EPR experiments indirectly reveal microwave induced generation of oxygen ion-radicals $O_2^-$ which are weakly bound to the zigzag edge of DCNO particles. These ion-radicals are fairly stable within that low temperature range, and coexist with antiferromagnetically coupled neighboring triplet oxygen molecules, physisorbed and stuck on the outermost nanographene shells of DCNOs and, thus, entrapped inside the nanopores of DCNO aggregates in the form of 2D and 3D clusters. These ion-radicals interact with edge spins forming new paramagnetic entities with $S = 1$. These entities irreversibly dissociate at $T > 60$ K, when both magnetic coupling and short-range ordering between physisorbed $O_2$ molecules become negligible against thermal fluctuations.

Both experiments and DFT calculations demonstrate that oxygen ion-radicals having $S = 1/2$ appear due to the transfer of π- electron from the nonbonding edge spin state on zigzag edges to triplet $O_2$. This charge transfer provides formation of weak bond between utmost atom of zigzag graphene edge and $O_2$ without further formation of covalent bonds and dissociation of oxygen molecule. Calculations for the various types of carbon substrate (mono- and bilayers, flat and corrugated, passivated by oxygen-containing and other realistic functional groups) demonstrate robustness of this effect because there are always unpaired π-electrons on zigzag edges. Varying of the shape of carbon substrate and functional groups on the edges provides oscillation of the charge transfer from substrate to oxygen molecule in the range 0.06~0.85 $e^-$ with the change of total magnetic moment of molecule to the values about 0.9 $\mu_B$ and binding energy in the range 40~720 meV. Despite the formation of this bond both experiment and theory demonstrate survival of molecular identity of oxygen molecule. Co-adsorption of other gases ($N_2$, $CO_2$ and $H_2O$) could change the values of charge transfer and binding energy of molecular oxygen but play the role only at low temperatures (below 40 K) due to weak bonds of these molecules with graphene edges. The presence of H- and COOH- terminated edge carbon sites with corrugated substrate morphology allows formation of ZE–$O_2^-$ paramagnetic complexes characterized by small (<50 meV) binding energies and also explains their irreversible dissociation above 50-60 K as revealed by EPR.

**Acknowledgements** The work is supported by the Ministry of Education and Science of the Russian Federation, Project N 16.1751.2014/K

**Table I.** Carbon-oxygen bond length (in Å), angle between oxygen-oxygen binuclear axis and graphene flat (in degrees), magnetic moment per oxygen molecule (in $\mu_B$), change of Mulliken population of oxygen and binding energies (in meV) for the various types of carbon substrates (see. Fig. 5) and functional groups on the edges.

| Number of layers | Buckling | Functional groups on edges | Distance | Angle | Magnetic moment | $\Delta e^-$ | $E_{binding}$ (meV) |
|---|---|---|---|---|---|---|---|
| Monolayer | Flat | -H (*) | 1.501 | 27.78 | 0.875 | 0.783 | 405 |
| | | -COOH | 1.496 | 27.64 | 0.898 | 0.770 | 370 |
| | | -OH | 1.462 | 30.27 | 0.881 | 0.814 | 718 |
| | Corrugated | -H | 1.604 | 25.23 | 0.948 | 0.764 | 72 |
| | | -COOH | 1.612 | 25.01 | 0.952 | 0.760 | 68 |
| | | -OH | 1.489 | 28.02 | 0.861 | 0.790 | 478 |
| Bilayer | Flat | -H | 1.586 | 26.57 | 0.924 | 0.771 | 160 |
| | | -COOH | 1.589 | 26.32 | 0.935 | 0.770 | 142 |
| | | -OH | 1.563 | 26.98 | 0.949 | 0.772 | 208 |
| | Corrugated | -H (**) | 1.632 | 24.63 | 0.963 | 0.768 | 46 |
| | | -COOH | 1.640 | 24.58 | 0.965 | 0.758 | 42 |
| | | -OH | 1.530 | 29.64 | 0.877 | 0.876 | 398 |

\* See Fig. 5b; \*\* See Fig. 5c

**Table II.** Binding energies and change of Mulliken population of oxygen molecule for the various types of co-adsorbents (see Fig. 6).

| co-adsorbent | $\Delta e^-$ | $E_{binding}$ (meV) | |
|---|---|---|---|
| | | oxygen | co-adsorbed gas |
| $O_2$ | 0.812 | 420 | 30 |
| $N_2$ | 0.862 | 1001 | 15 |
| $4N_2$ | 1.241 | 1486 | 14 |
| $H_2O$ | 0.899 | 1210 | 35 |
| $H_2O + N_2$ | 0.826 | 924 | 33 and 14 |
| $CO_2$ | 0.752 | 194 | 44 |
| $CO_2 + N_2$ | 0.772 | 233 | 42 and 15 |
| $H_2O + CO_2$ | 0.781 | 530 | 41 and 34 |
| $CO_2$ on the edge of corrugated bilayer | 0.757 | 44 | 30 |

# References

1. T. Enoki, Phys. Scr. Role of edges in the electronic and magnetic structures of nanographene.T146 (2012) 014008.

2. M. Fujita, K. Wakabayashi, K. Nakada, K. Kusakabe, Peculiar localized state at zigzag graphite edge. J. Phys. Soc. Jpn.65 (199601920.

3. K. Nakada, M. Fujita, G. Dresselhaus, M.S. Dresselhaus, Edge state in graphene ribbons: Nanometer size effect and edge shape dependence Phys. Rev. B 54 (1996) 17954.

4. T. Wassmann, A.P. Seitsonen, A.M. Saitta, M. Lazzeri, F. Mauri Clar's theory, π-electron distribution, and geometry of graphene nanoribbons. J. Am. Chem. Soc. 132 (2010) 3440.

5. Y. Kobayashi, K. Fukui, T. Enoki, K. Kusakabe, Y. Kaburagi, Observation of zigzag and armchair edges of graphite using scanning tunneling microscopy and spectroscopy. Phys. Rev. B 71 (2005) 193406.

6. Y. Kobayashi, K. Fukui, T. Enoki, K. Kusakabe, Edge state on hydrogen-terminated graphite edges investigated by scanning tunneling microscopy. Phys. Rev. B 73 (2006) 125415.

7. T. Enoki, Y. Kobayashi, K. Fukui, Electronic structures of graphene edges and nanographene. Inter. Rev. Phys. Chem. 26 (2007) 609.

8. S. Fujii, M. Ziatdinov, M. Ohtsuka, K. Kusakabe, M. Kiguchi, T. Enoki, Role of edge geometry and chemistry in the electronic properties of graphene nanostructures. Faraday Discus. 173 (2014) 173.

9. V.L.J. Joly, K.Takahara, K.Takai, K. Sugihara, T.Enoki, M.Koshino, H. Tanaka, Effect of electron localization on the edge-state spins in a disordered network of nanographene sheets. Phys. Rev. B. 81 (2010) 115408.

10. M. Ohtsuka, S. Fujii, M. Kiguchi, T. Enoki, Electronic state of oxidized nanographene edge with atomically sharp zigzag boundaries. ACS Nano 7(2013) 6868.

11. J. Takashiro, Y. Kudo, S. Kaneko, K. Takai, T. Ishii, T. Kyotani, T. Enoki, M. Kiguchi, Heat treatment effect on the electronic and magnetic structures of nanographene sheets investigated through electron spectroscopy and conductance measurements. Phys. Chem. Chem. Phys. 16 (2014) 7280.

12. D. Jiang, B.G. Sumpter, S. Dai, Unique chemical reactivity of a graphene nanoribbon's zigzag edge. J. Chem. Phys. 126 (2007) 134701.

13. O.E. Anderssonet al., Structure and electronic properties of graphite nanoparticles. Phys. Rev. B 58 (1998) 16387.

14. V. Yu. Osipov, et al., Magnetic and high resolution TEM studies of nanographite derived from nanodiamond. Carbon 44 (2006) 1225.

15. Y. Shibayama, H. Sato, T. Enoki, M. Endo, Disordered magnetism at the metal-insulator threshold in nano-graphite-based carbon materials. Phys. Rev. Lett. 84 (2000) 1744.

16. K. Takai, T. Suzuki, T. Enoki, H. Nishihara, T. Kyotani, Structure and magnetic properties of curved graphene networks and the effects of bromine and potassium adsorption. Phys. Rev. B 81 (2010) 205420.

17. V.Yu. Osipov, et al., Magnetic and EPR studies of edge-localized spin paramagnetism in multi-shell nanographites derived from nanodiamonds. Diam. Rel. Mater. 18 (2009) 220.

18. V.Yu. Osipov, et al., Interaction between edge-localized spins and molecular oxygen in multishell nanographites derived from nanodiamonds. Diam. Rel. Mater. 19 (2010) 492.

19. A. Tchernatinsky, S. Desai, G. U. Sumanasekera, C. S. Jayanthi, S. Y. Wu, B. Nagabhirava, B. Alphenaar, Adsorption of oxygen molecules on individual single-wall carbon nanotubes. J. Appl. Phys. 99 (2006) 034306.

20. Y. Sato, K. Takai, T. Enoki, Electrically controlled adsorption of oxygen in bilayer graphene devices. Nano Lett. 11 (2011) 3468.

21. R. Zacharia. Desorption of gases from graphitic and porous carbon surfaces. Berlin, Fritz-Haber-Institut der Max-Planck-Gesellschaft, PhD thesis, 2004.

22. N. Kobayashi, T. Enoki, C. Ishii, K. Kaneko, M. Endo, Gas adsorption effects on structural and electrical properties of activated carbon fibers. J. Chem. Phys. 109 (1998) 1983.

23. R. G. A. Veiga, R. H. Miwa, G. P. Srivastava, Quenching of local magnetic moment in oxygen adsorbed graphene nanoribbons. J. Chem. Phys. 128 (2008) 201101.

24. P. Giannozzi, R. Car, G. Scoles, Oxygen adsorption on graphite and nanotubes. J. Chem. Phys. 118 (2003) 1003.

25. A.I. Shames, V.Yu. Osipov, A.Ya. Vul', Y. Kaburagi, T. Hayashi, K. Takai, T. Enoki, Spin–spin interactions between π-electronic edge-localized spins and molecular oxygen in defective carbon nano-onions Carbon 61 (2013) 173.

26. K. Bogdanov, A. Fedorov, V. Osipov, T. Enoki, K. Takai, T. Hayashi, V. Ermakov, S. Moshkalev, A. Baranov, Annealing-induced structural changes of carbon onions: High-resolution transmission electron microscopy and Raman studies. Carbon 73 (2014) 78.

27. O. V. Yazyev, A guide to the design of electronic properties of graphene nanoribbons. Acc. Chem. Res. 46 (2013) 2319.

28. D. W. Boukhvalov, M. I. Katsnelson, Chemical functionalization of graphene. J. Phys. Condens. Matter 21 (2009) 344205.

29. B. Frank, R. Blume, A. Rinaldi, A. Trunschke, R. Schlögl, Oxygen insertion catalysis by *sp2* carbon. Angew. Chem. Int. Ed. 50 (2011) 10226.

30. D. W. Boukhvalov, X. Feng, K. Mullen, First-principles modeling of the polycyclic aromatic hydrocarbons reduction. J. Chem. Phys. C 115 (2011) 16001.

31. D. W. Boukhvalov, DFT modeling of the covalent functionalization of graphene: from ideal to realistic models. RSC Adv. 3 (2013) 7150.

32. J. M. Soler, E. Artacho, J. D. Gale, A. Garsia, J. Junquera, P. Orejon, D. Sanchez-Portal, The SIESTA method for ab initio order-N materials simulation. J. Phys. Condens. Matter 14 (2002) 2745.

33. J. P. Perdew, A. Zunger, Self-interaction correction to density-functional approximations for many-electron systems. Phys. Rev. B 23 (1981) 5048.

34. T. O. Wehling, K. S. Novoselov, S. V. Morozov, E. E. Vdovin, M. I. Katsnelson, A. K. Geim, A. I. Lichtenstein, Molecular doping of graphene. Nano. Lett. 8 (2008) 173.

35. E. Widenkvist, D.W. Boukhvalov, S. Rubino, S. Akhtar, J. Lu, R.A. Quinlan, M.I. Katsnelson, K. Leifer, H. Grennberg, U. Jansson, Mild sonochemical exfoliation of bromine-intercalated graphite: a new route towards graphene. J. Phys. D: Appl. Phys. 42 (2009) 112003.

36. D. W. Boukhvalov, Tuneable molecular doping of corrugated graphene. Surf. Sci. 604 (2010) 2190.

37. H. J. Monkhorst, J. D. Park, Special points for Brillouin-zone integrations. Phys. Rev. B 13 (1976) 5188.

38. O. N. Troullier, J. L. Martins, Efficient pseudopotentials for plane-wave calculations. Phys. Rev. B 43 (1991) 1993.

39. D. W. Boukhvalov, M. I. Katsnelson, Enhancement of chemical activity in corrugated graphene. J. Chem. Phys. C 13 (2009) 14176.

40. M. Yudasaka, T. Ichihashi, D. Kasuya, H. Kataura, S. Iijima, Structure changes of single-wall carbon nanotubes and single-wall carbon nanohorns caused by heat treatment. Carbon 41 (2003) 1273.

41. V. Yu. Osipov, A. E. Aleksenskiy, A. I. Shames, A. M. Panich, M. S. Shestakov, A. Ya. Vul', Infrared absorption study of surface functional groups providing chemical modification of nanodiamonds by divalent copper ion complexes. Diam. Relat. Mat. 20 (2011) 1234.

42. U. Kobler, R. Marx, Susceptibility study of physisorbed oxygen layers on graphite. Phys Rev B 35 (1987) 9809.

43. K. Kawamura, Y. Makishima, Y. Ochiai, Carbon – Sci. Tech. 2009, 2, 73.

44. Y.-W. Son, M. L. Cohen, S. G. Louie, Energy gaps in graphene nanoribbons. Phys. Rev. Lett. 97 (2006) 216803.

45. J. Cervenka, M. I. Katsnelson, C. F. J. Flipse, Room-temperature ferromagnetism in graphite driven by two-dimensional networks of point defects. Nat. Phys. 5 ( 2009) 840.

46. D. Martinez-Martin, M. Jaafar, R. Perez, J. Gomez-Herrero, A. Asenjo, Upper bound for the magnetic force gradient in graphite. Phys. Rev. Lett. 105 (2010) 257203.

47. CRC Handbook of Chemistry and Physics, 86th ed. (Ed.: D. R Lide), CRC Press, BocaRaton, FL, 2005.

48. I.-S. Byun, W. Kim, D. W. Boukhvalov, I. Hwang, J. W. Son, G. Oh, J. S. Choi, D. Yoon, H. Cheong, J. Baik, H.-J. Shin, H. W. Shiu, C.-H. Chen, Y.-W. Son, B. H. Park, Electrical control of nanoscale functionalization in graphene by the scanning probe technique. NPG Asia Mater. 6 (2014) e102.

49. T. O. Wehling, A. V. Balatsky, A. M. Tsvelik, M. I. Katsnelson, A.I. Lichtenstein, Midgap states in corrugated graphene: Ab initio calculations and effective field theory Europhys. Lett. 84 (2008) 17003.

50. H. Ulbricht, G. Moos, T. Hertel, Physisorption of molecular oxygen on single-wall carbon nanotube bundles and graphite. Phys. Rev. B. 66 (2002) 075404.

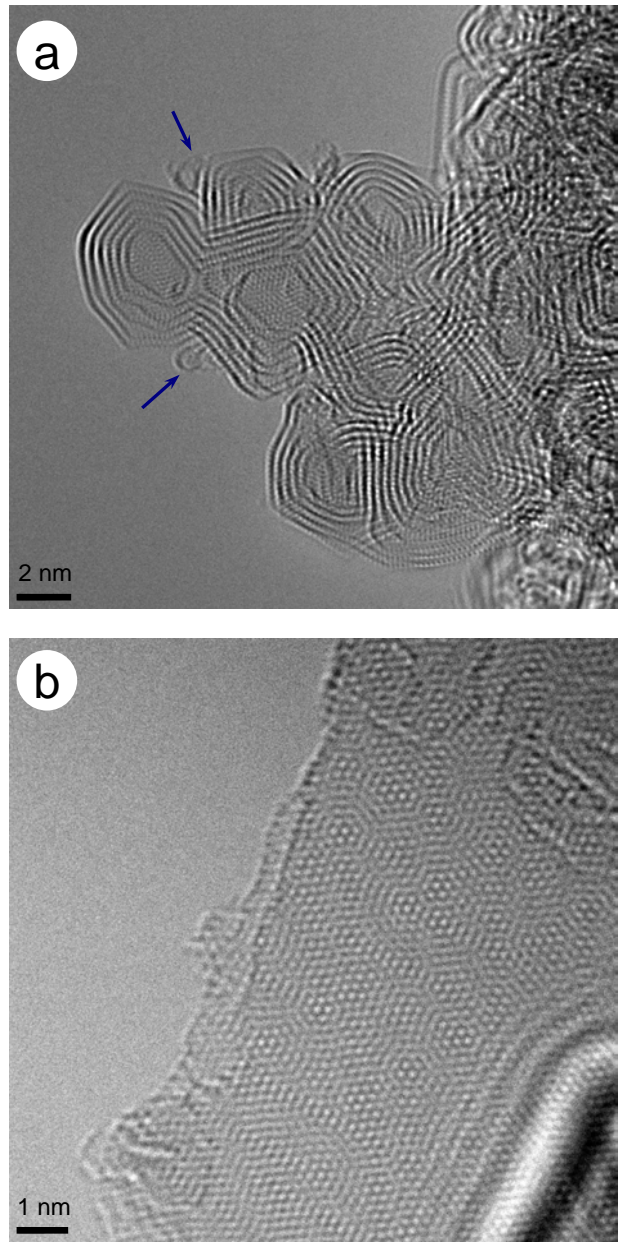

**Fig.1.** HRTEM images of characteristic morphological forms of the HT1950 sample: (a) general view of isolated and overlapped DCNO particles and their aggregates; (b) isolated few-layered graphene sheets sometimes occurred in the powder of DCNO particles in the minor quantity. Blue arrows in panel (a) indicate the position of splitted, bended and twisted nanographene strips appeared on the outer shells of carbon nano-onions. Well resolved honeycomb structures of nanographene facets and sheets together with the moiré patterns of slightly disorientated adjacent graphene sheets are clearly seen in fragments (a) and (b). Honeycomb structure of facet is well observable only for those complicated coalescent particle (see center of fragment (a)) consisting of two different polygon-like nuclei with common external graphene shells. Bars in each fragment correspond to the 2 nm (a) or 1 nm (b) scales.

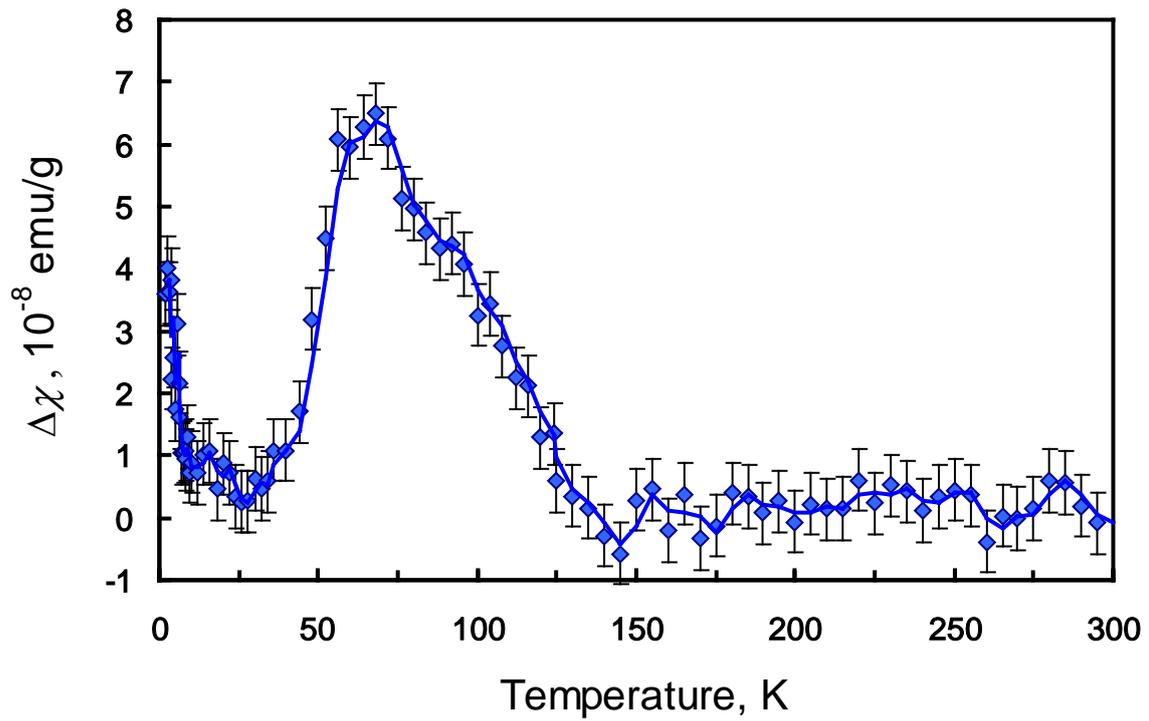

**Fig.2.** Temperature dependence of the difference between magnetic susceptibilities (Δχ) of air-physisorbed and physisorbed-gases-free HT1950 samples (measured in magnetic field 1 T). The error bar is determined by the technical problem related with inaccuracies in the subtraction process between the magnetization values of oxygen-containing and oxygen-free samples. The weight of HT1950 sample is 20.7 mg.

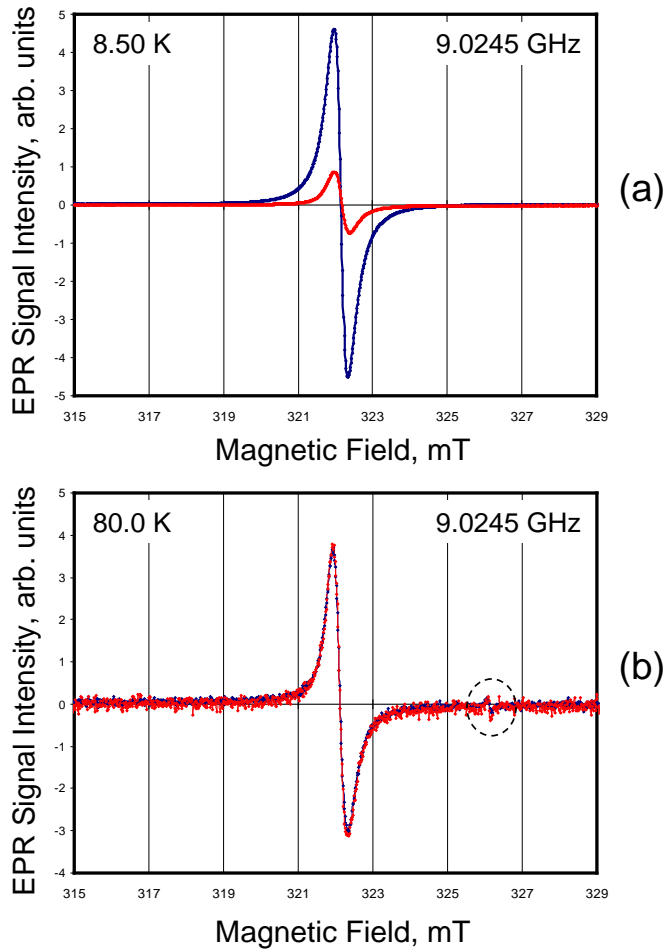

**Fig.3.** Narrow EPR signals in the DCNO HT1950-air sample measured at $T = 8.50$ K (a) and at elevated temperature $T = 80.0$ K (b) for two regimes of samples pretreatment prior the measurements at fixed $T$ and $P_{MW} = 100$ μW. Red curves traces – for pristine microwave unexposed sample initially kept at low ($T = 8.5$ K) or elevated temperatures ($T = 80$ K) in the cryostat during 20 min;   blue curves traces – for the same sample, but after its additional intense ~200 mW microwave irradiation during ~10 min at low stabilized temperature ($T = 8.5$ K) and subsequent abrupt decrease of microwave power down to 100 μW and stabilization or raise the temperature. Microwave power during the measurements – 100 μW. Both red and blue spectra are well coincided in the panel (b). Very weak EPR signal related with free oxygen molecules is also clearly seen in the circle in panel (b).

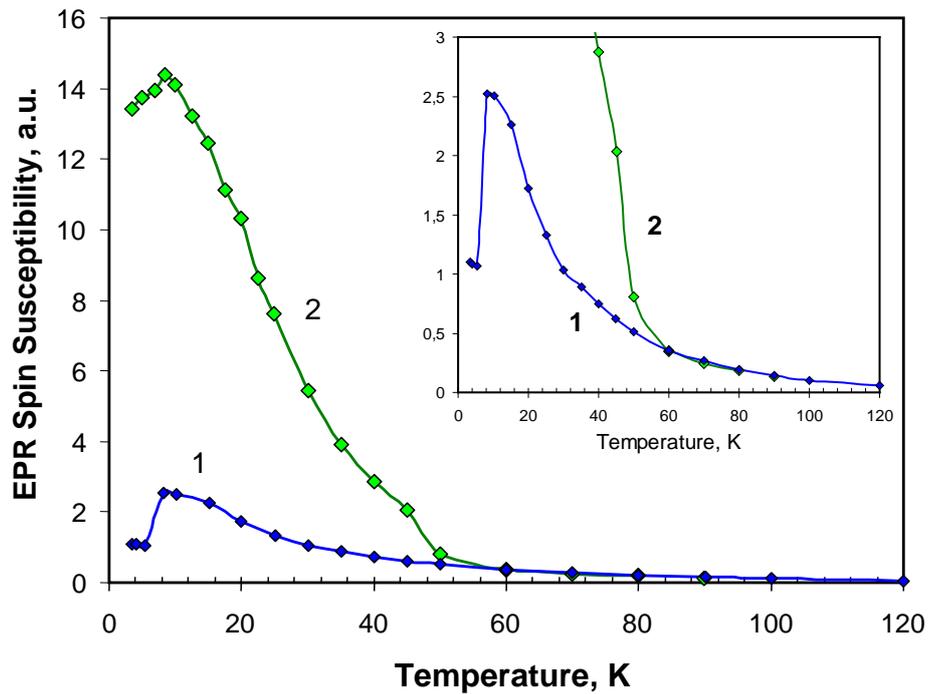

**Fig. 4.** Temperature dependence of EPR susceptibility in the air-physisorbed DCNO sample HT1950-air recorded at low $P_{MW}$ = 100 µW. Curve 1 (blue) – for the pristine sample cooled to 4 K at 100 µW never exposed to the microwave irradiation by higher $P_{MW}$; curve 2 (green) – after the short (10 min) exposure of the sample to $P_{MW}$ ~200 mW at 8.5 K and subsequent abrupt decrease of $P_{MW}$ down to 100 µW. At $T$ > 60 K both curves well coincide.

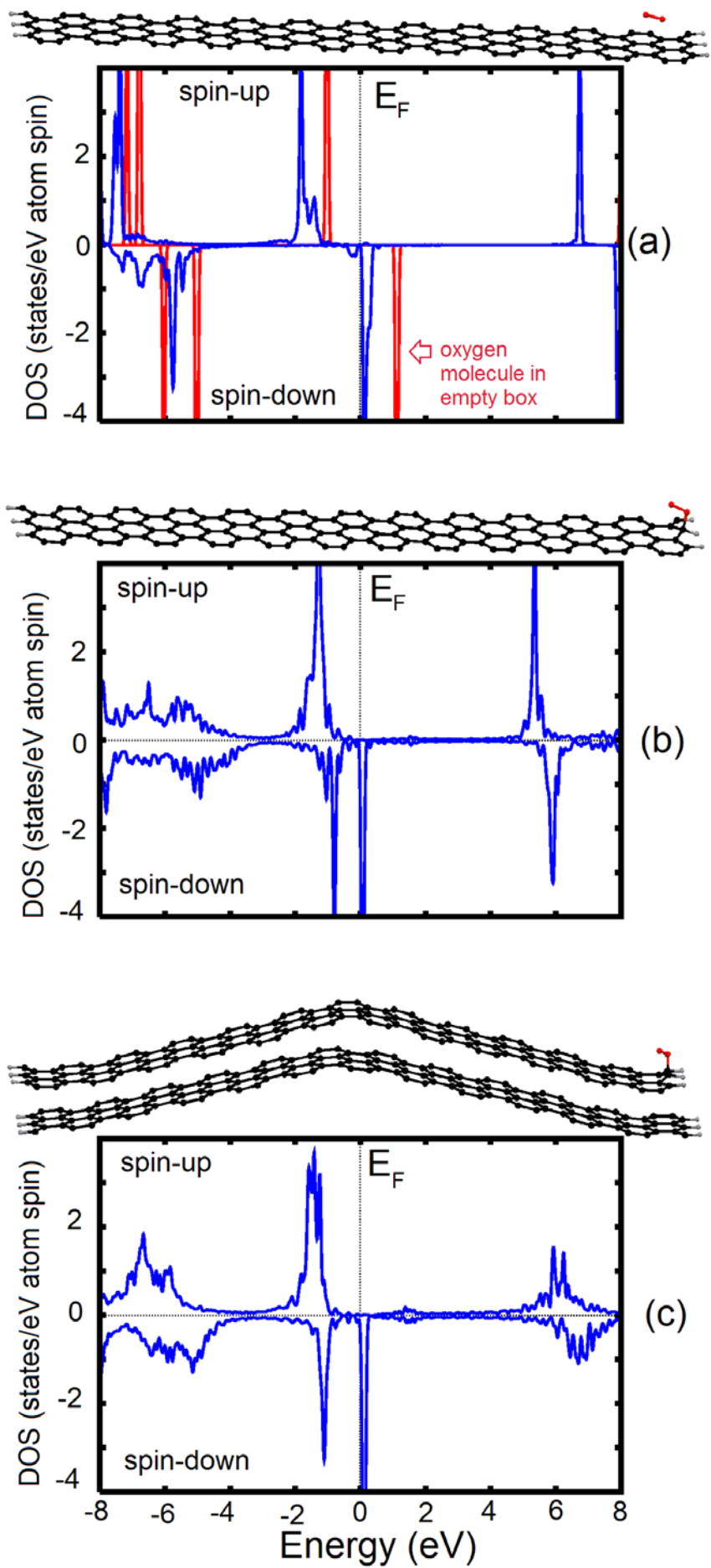

**Fig. 5.** Optimized atomic structures and spin polarized densities of states of oxygen molecule physisorbed on basal plane of graphene in vicinity of the edge (a), adsorbed on the edge of flat graphene monolayer (b) and corrugated graphene bilayer (c). For reference also reported density of states of oxygen molecule in empty box (red line on panel a).

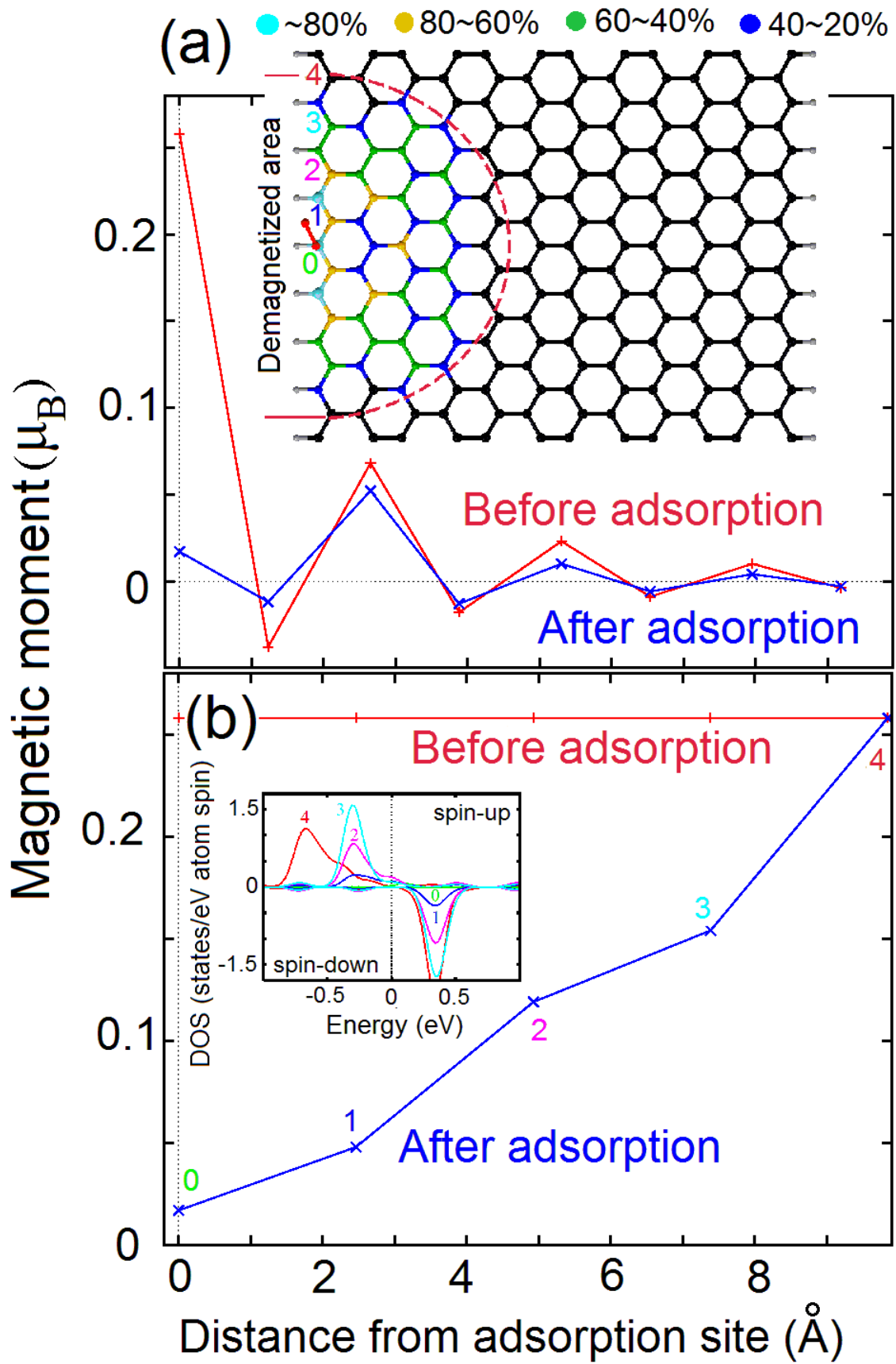

**Fig. 6.** The values of magnetic moments on the edge of graphene nanoribbon (see inset on panel a) before (red) and after (blue) adsorption of oxygen molecule on the edge carbon atom as function of distance from the atom connected with oxygen along nanoribbon (a) and across zigzag edge (b). On insets: (a) optimized atomic structure of graphene edge with carbon atoms indicated by different colors in respect to magnitude of decay of values of magnetic moments after adsorption of oxygen molecule and pointed by dashed red line area of charge harvesting (see discussion in text) and (b) spin-polarized densities of states for the edge carbon atoms marked by numbers on panel (b) and inset of panel (a).

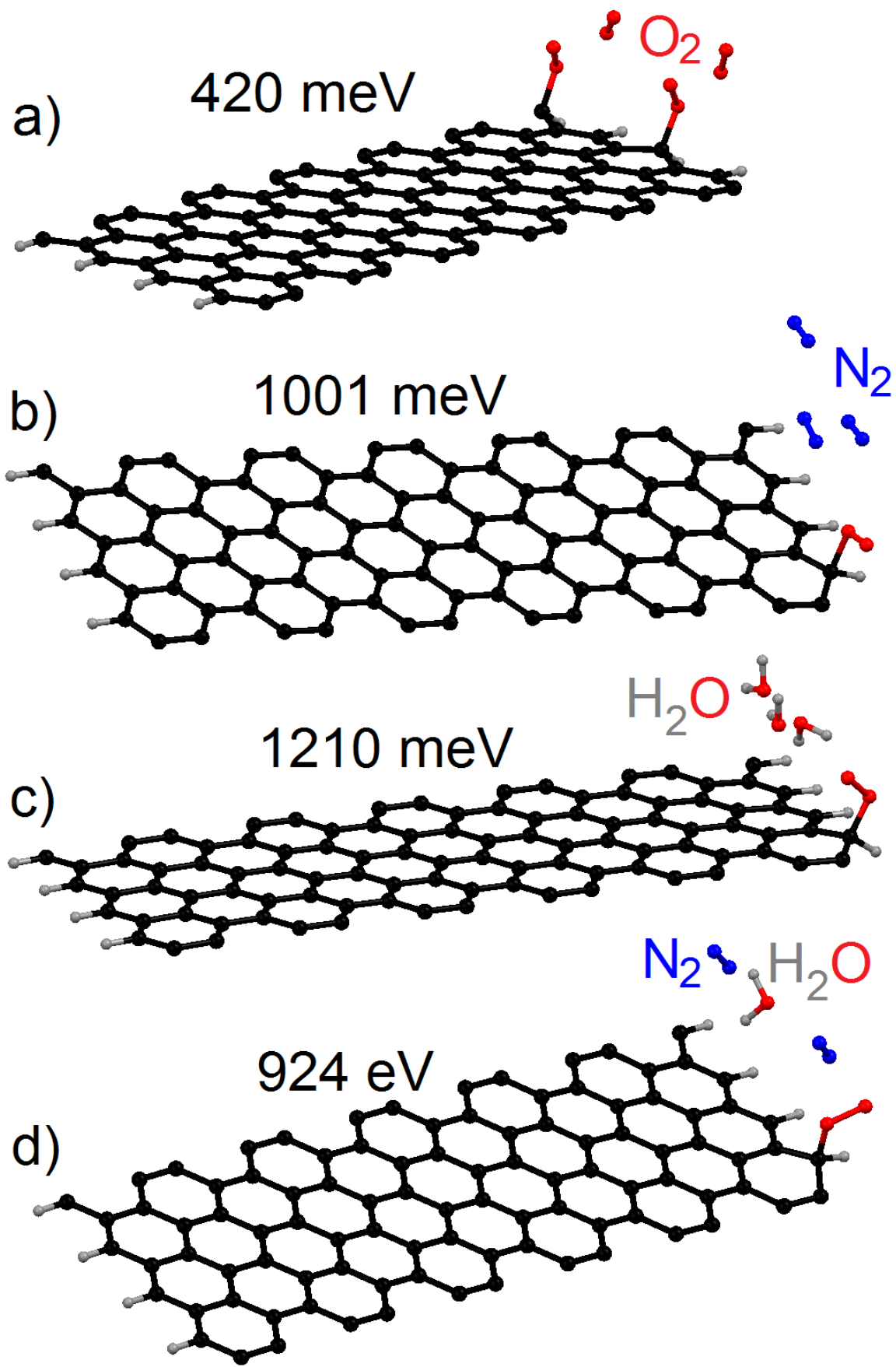

**Fig.7.** Optimized atomic structure and binding energies for adsorption of oxygen molecule over each edge atom (a), and co-adsorption of oxygen and nitrogen (b), water (c) and mix of water and nitrogen (d). The red, grey and blue balls denote oxygen, hydrogen and nitrogen atoms, respectively.

# Supplementary Information

## Charge transfer and weak bonding between molecular oxygen and graphene zigzag edges at low temperatures

D. W. Boukhvalov, V.Yu. Osipov, A.I. Shames, K. Takai, T. Hayashi, T. Enoki

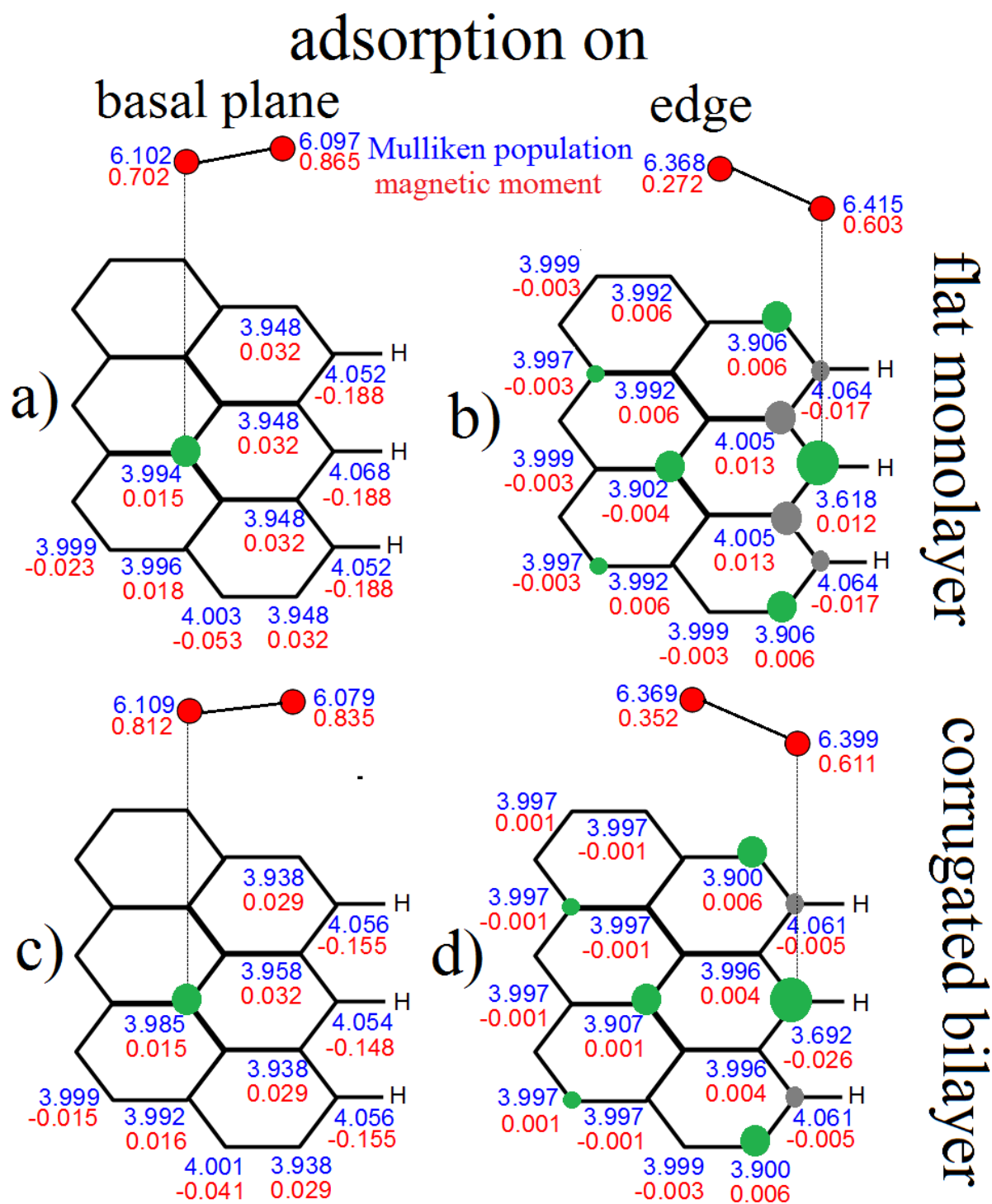

**Fig. S1.** A sketch of the graphene edges with adsorbed oxygen molecule. Mulliken populations ($e^-$, blue numbers) and magnetic moments ($\mu_B$, red numbers) for the adsorption of molecular oxygen on the basal plane in vicinity of edge (a, c) and on the utmost carbon atom of the edge (b, d) of flat monolayer (a, b) and corrugated bilayer graphene (c, d). The carbon atoms supplying the charge to oxygen molecule are indicated by green circles, carbon atoms with increased (above 4) electronic charge are indicated by gray circles (see discussion in text), the size of circles proportional to the change of Mulliken population of the atoms. (No optimized geometry for the position of oxygen molecule respecting the graphene edge is shown in this sketch).